# Features of the Galactic Cosmic Ray Anisotropy in the Solar Cycle 24 and Solar Minima 23/24 and 24/25

R. Modzelewska[1] ID,  K. Iskra[2] ID, W. Wozniak[3], M. Siluszyk[1] ID, M.V. Alania [4]

[1] Siedlce University of Natural Sciences and Humanities, Siedlce, Poland
[2] Polish Air Force University, Deblin, Poland
[3] Polish Gas Company, Warsaw, Poland
[4] Ivan Javakhishvili Tbilisi State University, Nodia Institute of Geophysics, Tbilisi, Georgia

Corresponding author: Renata Modzelewska  email: renatam@uph.edu.pl

**Abstract**
We study the role of drift effect in the temporal changes of the anisotropy of galactic cosmic rays (GCRs) and the influence of the sector structure of the heliospheric magnetic field on it. We analyze the GCRs anisotropy in the Solar Cycle 24 and solar minimum 23/24 with negative polarity ($qA<0$) for the period of 2007 - 2009 and near minimum 24/25 with positive polarity ($qA>0$) in 2017 - 2018 using data of global network of Neutron Monitors. We use the harmonic analyses method to calculate the radial and tangential components of the anisotropy of GCRs for different sectors ('+' corresponds to the positive and '-' to the negative directions) of the heliospheric magnetic field. We compare the analysis of GCRs anisotropy using different evaluations of the mean GCRs rigidity related to Neutron Monitor observations. Then the radial and tangential components are used for characterizing the GCRs modulation in the heliosphere. We show that in the solar minimum 23/24 in 2007-2009 when $qA<0$, the drift effect is not visibly evident in the changes of the radial component, *i.e.* the drift effect is found to produce a ≈4% change in the radial component of the GCRs anisotropy for 2007-2009. Hence the diffusion dominated model of GCRs transport is more acceptable in 2007-2009. In turn, near the solar minimum 24/25 in 2017-2018 when $qA>0$, the drift effect is evidently visible and produce a ≈40% change in the radial component of the GCRs anisotropy for 2017-2018. So in the period of 2017-2018 diffusion model with noticeably manifested drift is acceptable. The results of this work are in good agreement with the drift theory of GCRs modulation, according to which during negative (positive) polarity cycles, a drift stream of GCRs is directed toward (away from) the Sun, thus giving rise to a 22-year cycle variation of the radial GCRs anisotropy.

1. **Introduction**

Galactic cosmic rays (GCRs) measured at Earth are modulated in the heliosphere by expanding solar wind (SW). Modulation of GCRs is governed by the convection of SW, inward diffusion of GCRs on the irregularities of heliospheric magnetic field (HMF), drift on the regular HMF and adiabatic cooling of GCRs particles due to extension of SW. GCRs intensity detected near Earth contains the various long and short term quasi-periodic variations, see *e.g.* (Kudela and Sabbah, 2016; Chowdhury, Kudela, and Moon, 2016; Bazilevskaya *et al.,* 2014 and references therein). The most prominent periodic variations are 22 years, 11 years, 27 days and solar diurnal variation (≈ 24 hours). On the background of the long term solar modulation: 22 years and 11 years, connected with the global HMF and the solar activity (SA) cycle, respectively, the short term modulation effects sporadic (*e.g.* Forbush decreases) and recurrent (≈27 days) also take place.

This paper focuses on an anisotropic part of GCRs flux, reflected in the solar diurnal variation (≈24 hours wave), being a consequence of the equilibrium state between inward diffusion of GCRs in the HMF in the heliosphere and outward convection by the SW. The mechanism of the solar diurnal anisotropy was developed by (Ahluwalia and Dessler, 1962; Krymsky, 1964; Parker, 1964) based on the anisotropic diffusion-convection modulation theory of GCRs transport in the heliosphere, further developed by (Gleeson and Axford, 1967, 1968).

The anisotropy vector *A* is expressed in terms of the stream of GCRs and spatial density gradient, as (Gleeson, 1969):

$$A = -\frac{3S}{NV} = -\frac{3}{V}(K \cdot G - CU) \qquad (1)$$

where *N* is the GCRs density, $G = \frac{\nabla N}{N}$ is the density gradient, *S* is the stream of GCRs, *K* is the diffusion tensor representing the diffusion and drift effects of GCRs, *V* and *U* are speeds of the GCRs particles and of the SW, respectively, *C* is the Compton-Getting factor.

The components of the free-space anisotropy vector $A=[A_r, A_\theta, A_\phi]$ in spherical coordinates centered on the Sun are in the following form (Riker and Ahluwalia, 1987; Bieber and Chen, 1991; Alania *et al.,* 2001):

$$A_r^{\pm} = 3[CU - K_{rr}G_r^{\pm} \pm K_T G_\theta^{\pm} \sin\Psi + (K_{II} - K_\perp)G_\phi^{\pm} \sin\Psi\cos\Psi]/V \quad (2)$$

$$A_\theta^{\pm} = -3[\pm K_T G_r^{\pm} \sin\Psi + K_\perp G_\theta^{\pm} \pm K_T G_\phi^{\pm} \cos\Psi]/V \quad (3)$$

$$A_\phi^{\pm} = 3[(K_{II} - K_\perp)G_r^{\pm} \sin\Psi\cos\Psi \pm K_T G_\theta^{\pm} \cos\Psi - K_{\phi\phi}G_\phi^{\pm}]/V \quad (4)$$

where, $K_\perp$, $K_{II}$, and $K_T$ are perpendicular, parallel and drift diffusion coefficients of GCRs, respectively; $G_r^{\pm}$, $G_\theta^{\pm}$ and $G_\phi^{\pm}$ are the radial, heliolatitudinal and heliolongitudinal gradients of GCRs; sign (-) and (+) correspond to the HMF lines toward (negative) and away (positive) from the northern hemisphere of the Sun; $V$ and $U$ are speeds of the GCRs particles and of the SW, respectively; $\Psi$ is the angle between the HMF lines and the Earth-Sun line; $C=1.5$ for GCRs sensitive to neutron monitors (NMs) energy.

It can be seen from the system of Equations 2 - 4, that the GCRs anisotropy (recorded by NMs on the Earth) must be due to diffusion, convection and drift of GCRs particles in the regular HMF of the interplanetary space. It is, on average, 0.3-0.4%. At the same time, the fraction of the particle drift effect can reach 0.05-0.1% of the average anisotropy (Alania *et al.*, 1983, 1987, 2001; Alania, Bochorishvili, and Iskra, 2003). It is necessary to distinguish two types of drift effect in the GCRs anisotropy. The first type, due to the gradient and curvature of the HMF, can be identified in changes in the average anisotropy of GCRs in different periods of solar magnetic cycle *qA>0* and *qA<0* (for *qA>0*, the HMF lines come out from the northern hemisphere of the Sun, and for *qA<0*, they enter the northern hemisphere of the Sun). The second type of drift, due to the existence of local spatial gradients of the GCRs, is manifested in different sectors of the HMF. This component of drift is most notably linked to the heliospheric current sheet (HCS). The HCS is the surface where the polarity of the Sun's magnetic field changes from south to north and vice versa. The tilt angle (TA) is referred to the latitudinal extension (or waviness) of the HCS. We suppose that this drift effect can be manifested due to the sector structure of HMF for the minima epochs of SA, when the TA of the HCS is *<10-15* degrees, *i.e.* the Earth is located near the weakly wavy HCS. Revealing this type of drift may be difficult due to changes in SW speed and corotating interaction region (CIR) areas when the fast SW catches up slow SW (Richardson, 2018) disrupting the stability of the HMF sector structure. The absence of a significant azimuthal gradient in GCRs density when changing the HMF sign indicates a limited HCS thickness. This thickness should be many times smaller than the Larmor radius of GCRs particles with energy range of *1-10 GeV*. If the width of the HCS were of the order of the Larmor radius or greater, then the Earth, moving in orbit around the Sun, would pass areas with isotropic and anisotropic diffusion and we would observe a significant azimuthal GCRs density gradient (Alania and Dzhapiashvili, 1979).

However, in order to detect the first type of the drift effect in GCRs anisotropy, it is possible to average data over a long period of time (for example, over a period of ≈ 11 years). Whilst a clear identification of second type of drift effect is difficult on the background of changes in SW parameters for a short period of time comparable with the durations of the positive and negative sectors of the HMF.

Investigation of the role of drift effect in the temporal changes of the anisotropy of GCRs is very important from the point of view of the GCRs particle transport theory in the heliosphere (*e.g.* Siluszyk, Wawrzynczak, and Alania, 2011; Siluszyk, Iskra, and Alania, 2015). On the one hand, it allows to understand the behavior of GCRs particles transport in the heliosphere (Kota and Jokipii, 1983, 1998; Potgieter and Moraal, 1985; Burger and Potgieter, 1989; Potgieter, 1995; Alania *et al.*, 2001; Siluszyk *et al.*, 2018; Siluszyk, Iskra, and Alania, 2014; Iskra, Siluszyk, and Alania, 2015; Alania, Iskra, and Siluszyk, 2010; Iskra *et al.*, 2019). On the other hand, the clearly revealed drift effect in GCRs anisotropy allows to estimate the various parameters characterizing the SW and the diffusion of GCRs (Alania, Bochorishvili, and Iskra, 2003).

Recently, Dubey, Kumar, and Dubey (2018) analyzed the GCRs diurnal phase during 1986–2017 and confirmed its Hale cycle dependence (≈ 22 years connected with solar magnetic cycle). Park, Jung, and Evenson (2018) analyzed the diurnal variation comparing results of calculations for the pile-up method and harmonic analysis method. Oh, Yi, and Bieber (2010) found that the higher energy NMs contribute more to the 11 year phase variation of GCRs anisotropy due to the diffusion process. Sabbah (2013) reported that the diurnal phase of higher energy NMs shifts towards earliest hours; this is connected with outward convection by SW that increases the radial component of daily variation more than the azimuthal component.

The aim of this work is to continue research on the role of drift effect on temporary changes in GCRs anisotropy. We analyze in detail the anisotropy behavior in the Solar Cycle 24, especially during the solar minimum 23/24 in 2007-2009, when *qA<0* in comparison to the period 2017-2018 being near the solar minimum 24/25 when *qA>0*. The most important issue is: based on anisotropy analysis to determine the contribution of drift effect in GCRs modulation especially near solar minimum periods for Solar Cycle 24.

This paper is organized as follows: in Section 1 we have presented a short introduction and description of anisotropy of GCRs. Section 2 introduced data and methods used to calculate the GCRs anisotropy. In Section 3 we have presented experimental results and discussion. Section 4 concludes the paper.

## 2. Data and Methods

We use the hourly data of GCRs intensity for NMs with geomagnetic cut-off rigidities *Rc<5 GV*. For calculation of the GCRs anisotropy we follow the procedure from (Modzelewska and Alania, 2018; Ahluwalia *et al.*, 2015). Details of the NMs and parameters used for calculations are presented in Table 1.

| NM station | geo latitude $\lambda_{geo}[^0]$ | geo longitude $\Lambda_{geo}[^0]$ | asympt latitude $\lambda_{asym}[^0]$ | asympt longitude $\Lambda_{asym}[^0]$ | $R_c$ [GV] | CC min of SA | CC max of SA | CC intermediate | $R_{ef}$ [GV] | $R_m$ [GV] |
|---|---|---|---|---|---|---|---|---|---|---|
| Terre Adelie | -66.65 | 140.02 | -66.22 | 163.25 | 0.00 | 0.361 | 0.505 | 0.4330 | 12 | 15 |
| Nain | 56.55 | 298.30 | 23.79 | 337.01 | 0.30 | 0.677 | 0.610 | 0.6435 | 12 | 15 |
| Fort Smith | 60.00 | 248.00 | 30.60 | 266.64 | 0.30 | 0.568 | 0.505 | 0.5365 | 12 | 15 |
| Inuvik | 68.35 | 226.28 | 43.37 | 238.43 | 0.30 | 0.568 | 0.505 | 0.5365 | 12 | 15 |
| Thule | 76.60 | 291.20 | 67.36 | 320.64 | 0.30 | 0.234 | 0.210 | 0.2220 | 12 | 15 |
| Apatity | 67.57 | 33.40 | 38.04 | 68.58 | 0.65 | 0.592 | 0.529 | 0.5605 | 13 | 16 |
| Oulu | 65.05 | 25.47 | 32.40 | 62.92 | 0.80 | 0.619 | 0.560 | 0.5895 | 13 | 16 |
| Kerguelen | -49.35 | 70.25 | -8.74 | 81.77 | 1.14 | 0.733 | 0.682 | 0.7075 | 13 | 15 |
| Newark | 39.70 | 284.30 | -2.38 | 328.19 | 2.40 | 0.699 | 0.652 | 0.6755 | 14 | 17 |

**Table 1.** Details of each NM and parameters used for calculations.

First for calculations of the daily radial $A_r$ and tangential $A_\phi$ components of the diurnal variation of the GCRs intensity for each NM, we use the harmonic analyses method (*e.g.* Gubbins, 2004; Wolberg, 2006):

$$I(t) = \sum_{k=1}^{\infty}\left(A_r^k \cdot \cos\frac{2\pi kt}{T} + A_\phi^k \cdot \sin\frac{2\pi kt}{T}\right) = \sum_{k=1}^{\infty} A_k \cdot \sin\left(\frac{2\pi kt}{T} + \varphi_k\right) \quad (5)$$

where:

$$A_r^k = \frac{1}{p}\sum_{i=1}^{2p} y_i \cos\frac{\pi ki}{p}, \quad A_\phi^k = \frac{1}{p}\sum_{i=1}^{2p} y_i \sin\frac{\pi ki}{p}, \quad \varphi_k = \operatorname{arctg}\frac{A_\phi^k}{A_r^k}, \quad A_k = \sqrt{A_r^{2k} + A_\phi^{2k}},$$

where $y_i$ designates the hourly GCRs intensity data, *k* is the consecutive harmonic of the Fourier extension. The daily radial $A_r$ and tangential $A_\phi$ components of the diurnal variation of the GCRs intensity were calculated by means of normalized and detrended (excluding 25 hours trend) hourly data of the pressure corrected GCRs intensity as the first (*k=1, 2p=24* hours) harmonic of the Fourier extension.

We exclude from consideration the diurnal amplitudes *>0.7%* as an anomalous events related to the transient disturbances in the interplanetary space, generally connected with Forbush decreases and we do not take into account HMF sectors with duration less than 4 days. A number of the excluded days is less than *2-3%* from the total number of days used for analyses.

Next, in order to find local components of diurnal variation we rotate the vector by geographic longitude $\Lambda_{geo}$ of each NM station. Furthermore, we convert the radial $A_r$ and tangential $A_\phi$ components of the GCRs diurnal variation into the radial $A_r$ and azimuthal $A_\phi$ components of the GCRs anisotropy into the heliosphere (free space), (for details see Modzelewska and Alania, 2018), dividing the diurnal components by coupling coefficients (CC). The CCs represent the ratio of the diurnal variation to the corresponding amplitude of the anisotropy of cosmic rays in the heliosphere (Dorman, 1963; Yasue, Sakakibara, and Nagashima, 1982).

Calculated components of the diurnal variation of the GCRs intensity are corrected due to influence of the Earth magnetic field (Dorman *et al.*, 1972; Dorman, 2009), taking into account the asymptotic cone of acceptance characteristic for each NM station (Shea, Smart, and Mc Cracken, 1965; Shea *et al.*, 1967) for the rigidity to which NMs respond. The asymptotic cones of acceptance for each NM station were calculated using the FORTRAN TJ2000 program developed by Shea *et al.* (1967). Geomagnetic field coefficients were taken for the epoch of 2015 of the International Geomagnetic Reference Field. It is done by the rotation of the corresponding angle, called geomagnetic bending (GB), $\varphi_{GB} = \Lambda_{asym} - \Lambda_{geo}$ characteristic for each NM station, $\Lambda_{asym}$ and $\Lambda_{geo}$ are the asymptotic and geographic longitude. The radial and the tangential components of the diurnal variation of the GCRs intensity corrected to the Earth magnetic field influence are expressed (*e.g.* Ygbuhay, 2015):

$$A_r = a_r \cdot \cos(\varphi_{GB}) - a_\phi \cdot \sin(\varphi_{GB}) \qquad (6)$$
$$A_\phi = a_r \cdot \sin(\varphi_{GB}) + a_\phi \cdot \cos(\varphi_{GB}) \qquad (7)$$
$$\varphi_{free\ space} = \varphi_{local\ phase} + \varphi_{GB} \qquad (8).$$

Finally we correct the GCRs anisotropy for Compton-Getting (CG) effect, due to Earth orbital motion (Ahluwalia and Ericksen, 1970; Hall, Duldig, and Humble, 1996). CG correction is a vector with amplitude 0.045 cos $\lambda_{asym}$ and phase 6h, where $\lambda_{asym}$ is the asymptotic latitude of each NM.

The GCRs intensity measured by ground-based NMs is based on the concept of an integrated GCRs flux above the magnetic rigidity threshold that depends on the geomagnetic cut-off typical for each NM. In this case there arises a question what is the real energy of the measurement. It is natural to introduce the "effective energy/rigidity" typical for GCRs detector, here NM. In the literature this term has different meanings: fixed energy/rigidity ≈ *10 GV* for a NM, (Belov, 2000), median energy (Ahulwalia and Dorman, 1997; Jämsén *et al.*, 2007), maximum of the differential response function (McCracken *et al.*, 2004), or the integral effective energy (Alanko *et al.*, 2003). Unfortunately the above mentioned effective energy concepts are not constant but are changing through the solar activity cycle. Recently, Gil *et al.* (2017) proposed a model of the effective energy in which the variability of the GCRs flux at this energy is directly proportional to the detector's count rate, so that the percentage variability of the detector's count rate is equal to that of the GCRs flux at this energy, for details see Asvestari *et al.* (2017).

In this paper we compare the analysis of the GCRs anisotropy using this two approaches: the median rigidity $R_m$ of NM response (Ahluwalia *et al.*, 2015; Ahluwalia and Fikani, 2007) and the effective rigidity $R_{ef}$ characteristic for each NM proposed by (Gil *et al.*, 2017; Asvestari *et al.*, 2017).

3. **Experimental Results and Discussion**

Solar Cycle 24 was much less active than, for example, Solar Cycle 23, being rather similar to cycles at the beginning of the previous century. During this relatively weak cycle not many impulsive events happened on the Sun, but we could observe the prolonged solar minimum 23/24 with good established regular structure of the HMF. The current Solar Cycle 24 has now more or less finished its descending phase. At the beginning of 2018 (≈ April), the Sun showed indications of the reverse magnetic polarity sunspot announcing the Solar Cycle 25 (Phillips, 2018). The poleward reversed polarity sunspots suggest that a transition to Solar Cycle 25 is in progress and indicating that we may be in- or near the minimum between Solar Cycles 24 and 25. So it is of special importance to study the features of the GCRs anisotropy characteristics through the almost whole Solar Cycle 24 with special importance of the comparison of the drift effect in GCRs anisotropy during solar minimum 23/24 and near the solar minimum 24/25 with different polarity periods of solar magnetic cycle.

Here we present the time variation of the GCRs anisotropy for Oulu NM (as an example) through the almost whole Solar Cycle 24, 2007-2018. We compare the values of the amplitude, phase, radial and azimuthal components of the GCRs anisotropy using two approaches: median $R_m$ (Ahluwalia *et al.*, 2015) and effective $R_{ef}$ (Gil *et al.*, 2017) rigidity of response. The errors for annual values are estimated as the standard deviation from monthly data. The calculated results of the GCRs anisotropy for effective and median rigidity are in good agreement, the difference is in scope of the error bars. Figure 1 presents the time lines of the annual average of the amplitude (Figure 1a), phase (Figure 1b), radial (Figure 1c) and azimuthal (Figure 1d) components of the GCRs anisotropy for 2007-2018. The amplitude of the GCRs anisotropy displays the 11-year variation with systematic decrease of the values near the minima of SA (Figure 1a). The phase of the GCRs anisotropy shows the undoubted indication of shifting towards earlier hours near the solar minimum 24/25 (Figure 1b). Figure 1c demonstrates the change of $A_r$ component from positive to negative values close the maximum of Solar Cycle nearby to the reversal of the global HMF.

As far the differences of the GCRs anisotropy parameters calculated for these two approaches: median and effective rigidity are rather negligible, we decided to use further in this paper only GCRs anisotropy using recent $R_{ef}$ methodology following Gil *et al.* (2017).

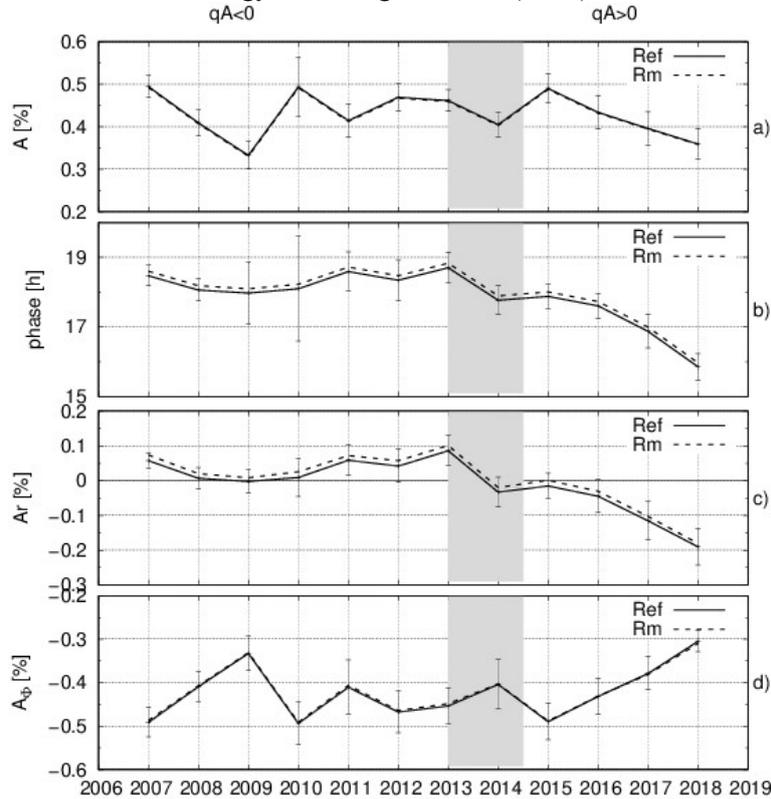

**Figures 1abcd.** Temporal changes of the annual average amplitude $A$ (a), phase (b), radial $A_r$ (c) and azimuthal $A_\phi$ (d) components of the GCRs anisotropy by Oulu NM for the almost whole Solar Cycle 24, 2007-2018. Shadow area designates the solar maximum 2013-2014.

The aim of this work is to continue research on the role of drift effect on temporary changes in GCRs anisotropy. The most important issue is: based on the anisotropy analysis to determine the contribution of drift effect in GCRs modulation especially near solar minimum periods for Solar Cycle 24. In this purpose components $A_r$ and $A_\phi$ of the anisotropy of GCRs were calculated using the harmonic analyses of hourly data of every separate NM for periods of two minima: solar minimum 23/24: 2007-2009, when $qA<0$ and near solar minimum 24/25: 2017-2018, when $qA>0$.

In Figures 2abc are presented average GCRs anisotropy vectors from following NMs: Apatity, Oulu, Newark, Thule, Fort Smith, Inuvik, Kerguelen, Terre Adelie and Nain (see Table 1) for different sectors of the HMF ('+' corresponds to the positive and '-' to the negative sectors) for the solar minimum 23/24 when *qA<0*: 2007 (Figure 2a), 2008 (Figure 2b), and 2009 (Figure 2c). In 2007 data for Terre Adelie and Kerguelen NMs and in 2008 for Terre Adelie NM are not available.

In Figures 3ab are presented average vectors of the GCRs anisotropy for different sectors of the HMF for the period near the solar minimum 24/25 when *qA>0*: 2017 (Figure 3a) and 2018 (Figure 3b). Here we used data of the same NMs as in Figure 2.

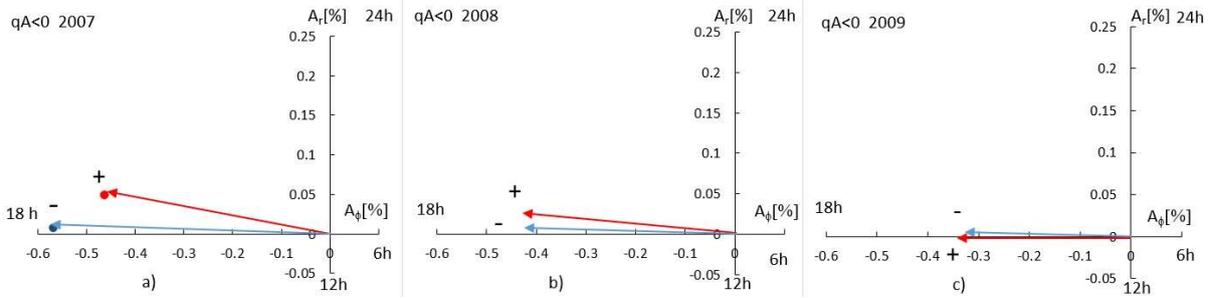

**Figures 2abc.** Harmonic diagram of the GCRs anisotropy vectors for positive (+) and negative (-) sectors of HMF for solar minimum 23/24 when *qA<0*: 2007 (Figure 2a), 2008 (Figure 2b) and 2009 (Figure 2c). On the vertical and horizontal axes are the radial and azimuthal components, in %.

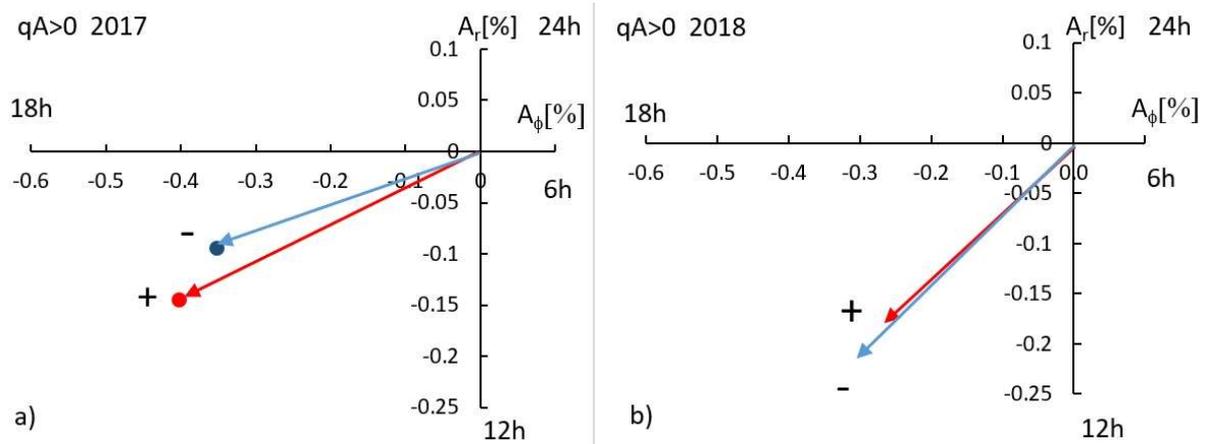

**Figures 3ab.** The same as in Figure 2 but near solar minimum 24/25 when *qA>0*: 2017 (Figure 3a) and 2018 (Figure 3b).

Table 2 presents the radial and azimuthal components, magnitude of the GCRs anisotropy and the phase relative to 18h. Uncertainty of this data is estimated as the Poisson statistics using standard deviations for separate NMs devided by $\sqrt{n}$, where *n* is the number of NMs used in averaging for each period. As far the drift stream is evidently pronounced in the radial component we present in Table 2 also the ratios of the absolute value of the radial component to the value of the magnitude of the GCRs anisotropy vector. All quantities are for different HMF sectors in 2007, 2008, 2009 and averaged values for 2007-2009, when *qA<0* and the same in 2017, 2018 and averaged for the period 2017-2018, when *qA>0*.

| Year | Ar+ [%] | Ar- [%] | Aφ+ [%] | Aφ- [%] | A+ [%] | A- [%] | \|Ar+\|/\|A+\| [%] | \|Ar-\|/\|A-\| [%] | phase+ [h:mm] | phase- [h:mm] |
|---|---|---|---|---|---|---|---|---|---|---|
| 2007 | 0.049±0.005 | 0.007±0.005 | -0.462±0.005 | -0.568±0.005 | 0.465±0.004 | 0.568±0.004 | 10.6±1.1 | 1.3±0.8 | 00:24 | 00:03 |
| 2008 | 0.025±0.005 | 0.010±0.005 | -0.424±0.005 | -0.422±0.005 | 0.425±0.004 | 0.423±0.004 | 6.0±1.1 | 2.3±1.2 | 00:14 | 00:05 |
| 2009 | -0.003±0.004 | 0.003±0.004 | -0.346±0.005 | -0.336±0.004 | 0.346±0.003 | 0.336±0.003 | 0.8±1.3 | 1.0±1.3 | -00:02 | 00:02 |
| 2007-2009 | 0.024±0.005 | 0.007±0.005 | -0.411±0.005 | -0.442±0.005 | 0.411±0.004 | 0.442±0.004 | 5.8±2.0 | 1.5±1.3 | 00:13 | 00:04 |
| 2017 | -0.145±0.005 | -0.094±0.007 | -0.403±0.004 | -0.353±0.007 | 0.429±0.004 | 0.365±0.005 | 33.8±0.9 | 25.8±1.7 | -01:19 | -01:00 |
| 2018 | -0.183±0.006 | -0.217±0.005 | -0.268±0.006 | -0.310±0.004 | 0.324±0.005 | 0.379±0.005 | 56.3±1.1 | 57.5±0.6 | -02:17 | -02:20 |
| 2017-2018 | -0.164±0.006 | -0.156±0.007 | -0.336±0.006 | -0.331±0.007 | 0.376±0.005 | 0.366±0.005 | 43.8±0.4 | 42.6±3.7 | -01:44 | -01:41 |

**Table 2** The values of the radial and azimuthal components, magnitude of the GCRs anisotropy, the phase relative to 18h and the ratios of the absolute value of the radial component to the value of the magnitude of the GCRs anisotropy vector for different HMF sectors in 2007, 2008, 2009 and averaged values for 2007-2009, when *qA<0* and the same in 2017, 2018 and averaged for the period 2017-2018, when *qA>0*.

From Figures 2abc and Table 2, we can see that all vectors of the GCRs anisotropy for both positive and negative sectors of HMF are slightly shifted towards later hours in relation to 18 hour for 2007-2009, when *qA<0*. In 2007 the shifting is 24 min for positive sectors and 3 min for negative sectors, in 2008 - 14 min and 5 min all after 18h, respectively. These values are very small and are in scope of the error bars. In 2007 and 2008 the vectors of the GCRs anisotropy for positive sectors are shifted towards the later hours than the vector for negative sectors. The ratios of the absolute value of the radial component to the value of the total anisotropy vector are 10.6±1.1% and 1.3±0.8% in 2007 for positive and negative sectors and 6.0±1.1% and 2.3±1.2% in 2008, respectively.

There is an unusual behavior of the vector of the GCRs anisotropy in 2009. The vector of the GCRs anisotropy for positive sectors practically points to 18 hours (phase + 2 min before 18h, phase – 2 min after 18h). In 2009 the radial components of the anisotropy vectors for different sectors are almost zero ($A_r$+ is -0.003±0.004% towards 12h and $A_r$- is 0.003±0.004% towards 24h). The ratios of the absolute value of the radial component to the value of the total anisotropy vector in 2009 are 0.8±1.3% and 1.0±1.3% for positive and negative sectors, respectively. These quantities are practically the same within the error limits.

This means that in the above periods diffusion is responsible for the propagation of cosmic radiation in the heliosphere, while the drift of cosmic ray particles due to the gradient and curvature is insignificant and in 2009 it hardly exists. One can see also, that there are small differences between the amplitudes and phases of the GCRs anisotropy vectors for different sectors of the HMF in 2008 and 2009. Moreover the effect of drift in the anisotropy of GCRs due to the sector structure of the HMF during the minima epochs of SA is generally caused by the drift on the HCS and is mostly pronounced in the differences of azimuthal components between positive and negative sectors in 2007.

     From Figures 3ab and Table 2 one can see that all vectors of the GCRs anisotropy for both positive and negative sectors of HMF are shifted towards earlier hours compared to 18 hour in 2017-2018, when *qA>0*. In 2017 (Figure 3a) the magnitude of the vector of the GCRs anisotropy for negative sector of HMF is less (A-=0.365±0.005%) than the magnitude of the vector for positive sector (A+=0.429±0.004%). We have the reverse situation in 2018 (Figure 3b) (A+=0.324±0.005%), is less than (A-=0.379±0.005%). The anisotropy vector for negative sector is shifted towards the later hours in 2017, at the same time we observe no significant difference in the phase for both sectors in 2018. The ratios of the absolute value of the radial component to the value of the total anisotropy vector are 33.8±0.9% and 25.8±1.7% in 2017 and in 2018 the ratios are even higher being 56.3±1.1% and 57.5±0.6% for positive and negative sectors, respectively. The ratios for various sectors in 2018 are practically the same within the error bars. The effect of drift in the anisotropy of GCRs due to the sector structure of the HMF is mostly pronounced in the differences of azimuthal components between positive and negative sectors in 2017 and 2018.

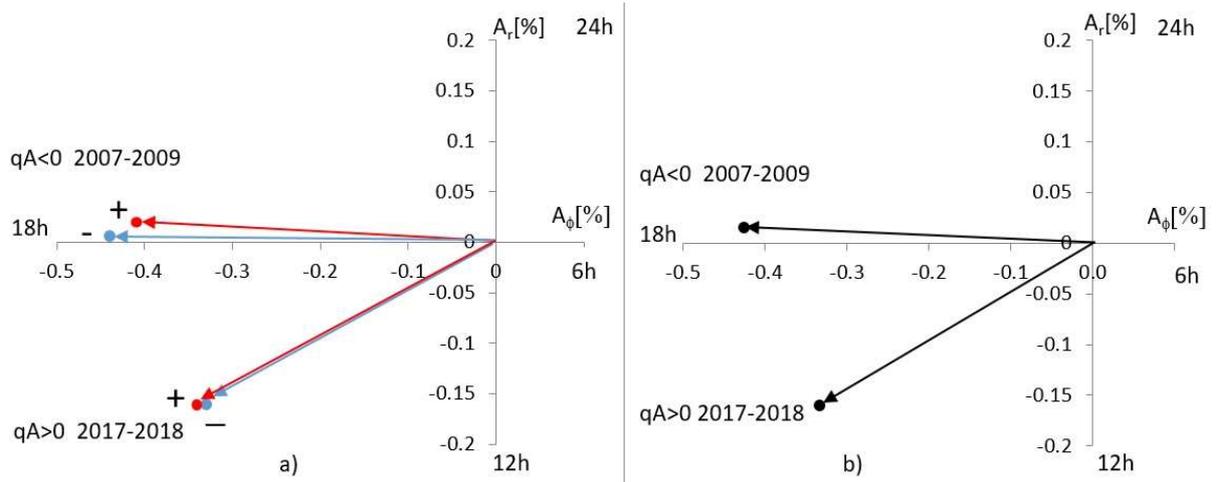

**Figures 4ab.**
(a) Harmonic diagram of the average GCRs anisotropy for positive (+) and negative (-) sectors of HMF for the periods 2007-2009 when *qA<0* and 2017-2018 when *qA>0*; (b) harmonic diagram of the average over sectors the GCRs anisotropy for 2007-2009 when *qA<0* and 2017-2018 when *qA>0* for the global HMF.

| Year | $A_r$[%] | $A_\phi$[%] | $A_{av}$[%] | $|A_r|/A_{av}$ [%] | phase [h:mm] |
|---|---|---|---|---|---|
| 2007-2009 | 0.015±0.036 | -0.426±0.038 | 0.427±0.029 | 3.6±2.4 | 00:14 |
| 2017-2018 | -0.160±0.053 | -0.333±0.049 | 0.370±0.042 | 43.3±11.6 | -01:09 |
| 2007-2018 | -0.012±0.023 | -0.422±0.018 | 0.422±0.015 | 2.8±3.8 | -00:06 |

**Table 3** The values of the radial and azimuthal components, average magnitude of the GCRs anisotropy, the ratio of the absolute radial component to the average magnitude of the GCRs anisotropy and the phase relative to 18h for 2007-2009, when *qA<0* and the same for the period 2017-2018, when *qA>0* and for the almost whole Solar Cycle 24, 2007-2018.

Figure 4a and Table 2 show the average GCRs anisotropy vectors for different sectors of the HMF averaged over the periods 2007 - 2009, when *qA<0* and 2017 - 2018, when *qA>0*. One can see that in 2007-2009 the magnitude of the vector for positive (+) sectors of HMF is a bit less (*A+*=0.411±0.004%) than the magnitude of the vector for negative sectors (*A-*=0.442±0.004%). Phases of the vectors A+ and A- are directed 13 min and 4 min respectively after 18h in 2007-2009. Whereas in 2017-2018 there is no noticeable difference between positive and negative sectors; the magnitude of the anisotropy vectors is ≈ 0.37% for both sectors, and the phase is shifted ≈ 1:40h before 18h. The ratios of the absolute value of the radial component to the value of the total anisotropy vector are for the whole solar minimum 23/24 in 2007-2009 when *qA<0*: 5.8±2.0% and 1.5±1.3% for (+) and (-) sectors, respectively and near the solar minimum 24/25 in 2017-2018 when *qA>0* the ratios are 43.8±0.4% and 42.6±3.7% for (+) and (-) sectors, respectively and practically are the same within the error bars.

Having calculated anisotropy vectors for different sectors, we average the vectors of the GCRs anisotropy over different sectors but for the same directions of the global HMF. Figure 4b and Table 3 show averaged over HMF sectors the anisotropy vectors for different directions of the global HMF *i.e.* for 2007-2009, when *qA<0* and 2017-2018, when *qA>0*.

Summarizing, in the solar minimum 23/24 in 2007-2009 when *qA<0*, the drift effect is not pronounced visibly in the changes of the $A_r$ component being almost 0 ($A_r$=0.02±0.04%), determining the phase shifted 14 min after 18h, *i.e.* for this solar minimum the diffusion dominated model of GCRs transport is more acceptable in 2007-2009. In turn, near the solar minimum 24/25 in 2017-2018 when *qA>0*, the radial component $A_r$ is -0.16±0.05% and the phase is shifted 1:09h prior 18h. So, the drift effect is clearly visible. So in the period of 2017-2018 the diffusion model with clearly manifested drift is acceptable. For the whole solar minimum 23/24 in 2007-2009 when *qA<0* the ratio of the absolute value of the radial component to the value of the total anisotropy vector is 3.6±2.4%, whereas near the solar minimum 24/25 in 2017-2018 when *qA>0* this ratio is 43.3±11.6%.

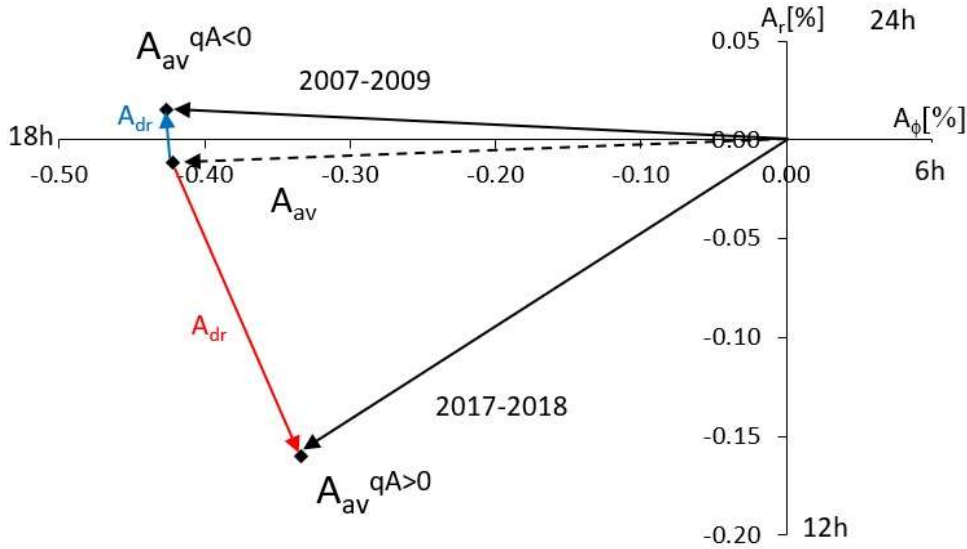

**Figure 5.** Harmonic diagram of the average vector of the GCRs anisotropy $A_{av}^{qA<0}$ for 2007-2009, $A_{av}^{qA>0}$ for 2017-2018 and $A_{av}$ for the almost whole Solar Cycle 24 in 2007-2018 (dashed line). Figure 5 presents the drift anisotropy vectors $A_{dr}$ for *qA<0* (blue) and $A_{dr}$ for *qA>0* (red).

In Figure 5 are presented the average vector of the GCRs anisotropy $A_{av}^{qA<0}$ for 2007-2009, $A_{av}^{qA>0}$ for 2017-2018 and $A_{av}$ for the almost whole Solar Cycle 24 in 2007-2018 (dashed line). Figure 5 also presents the drift vectors $A_{dr}$ for *qA<0* (blue) calculated as ($A_{av}^{qA<0}$ - $A_{av}$) and $A_{dr}$ for *qA>0* (red) calculated as ($A_{av}^{qA>0}$ - $A_{av}$).

One can see that the drift anisotropy vector $A_{dr}$ (red colour in Figure 5) for the *qA>0* polarity is preferentially directed to the Sun (to the direction of 12h in the harmonic diagram), and for the *qA<0* polarity the drift anisotropy vector $A_{dr}$ (blue colour in Figure 5) is directed out of the Sun. These empirical outcome are in good correspondence with the drift theory of modulation of GCRs (Jokipii, Levy, and Hubbard, 1977). Based on that theory in the *qA>0* cycle a drift stream of GCRs caused by the gradient and curvature of the HMF is preferentially coming from the polar regions to the helioequatorial region and is directed away from the Sun. For the *qA<0* cycle exists the opposite direction of the drift stream of GCRs.

Table 4 presents the radial and azimuthal components and magnitude of the drift GCRs anisotropy vector $A_{dr}$ and the ratios of the absolute value of the $A^r_{dr}$ components to the total magnitude of $A_{dr}$ for 2007-2009, when *qA<0* and 2017-2018, when *qA>0*. In the solar minimum 23/24 in 2007-2009, when *qA<0*, the ratio $A^r_{dr}/A_{dr}$ is ≈ 99%. In turn, near the solar minimum 24/25 in 2017-2018 when *qA>0*, the ratio $A^r_{dr}/A_{dr}$ reaches ≈ 86%.

Thus, the drift vectors $A_{dr}$ of the GCRs anisotropy (Figure 5) for *qA>0* and *qA<0* polarities basically contain the radial component, the contribution of $A^r_{dr}$ in total $A_{dr}$ is more than ≈ 86% for *qA>0* and almost ≈ 99% in *qA<0*. The difference between anisotropy *A* for *qA>0* and *qA<0* polarities is produced by the drift of GCRs due to the curvature and gradient of the regular HMF and is the source of the long term variation of the anisotropy *A* of GCRs (Forbush, 1969; Ahluwalia, 1988).

| Year | $A^r_{dr}$[%] | $A^\phi_{dr}$[%] | $A_{dr}$[%] | $|A^r_{dr}|/A_{dr}$ [%] | $|A^\phi_{dr}|/A_{dr}$ [%] |
|---|---|---|---|---|---|
| 2007-2009 | 0.027±0.04 | -0.004±0.04 | 0.028±0.04 | 98.9 | 14.7 |
| 2017-2018 | -0.148±0.05 | 0.089±0.05 | 0.173±0.05 | 85.7 | 51.5 |

**Table 4.** The radial, azimuthal components and magnitude of the drift GCRs anisotropy $A_{dr}$ and the ratios of the absolute value of the components of $A^r_{dr}$ to the total magnitude of $A_{dr}$ for 2007-2009, when $qA<0$ and 2017-2018, when $qA>0$.

Summarizing, in the solar minimum 23/24 in 2007-2009, when $qA<0$, the drift effect is not pronounced visibly, the ratio $|A_r|/A_{av}$ is only ≈ 4%. Hence the diffusion dominated model of GCRs transport is more acceptable in 2007-2009. In turn, near the solar minimum 24/25 in 2017-2018 when $qA>0$, the drift effect is clearly visible and the ratio $|A_r|/A_{av}$ reaches ≈ 40%. So in the period of 2017-2018 the diffusion model with clearly manifested drift is acceptable. The above mentioned statements are in agreement with other authors. According to (Moraal and Stoker, 2010; Mewaldt *et al.*, 2010) the solar minimum 23/24 was diffusion dominated with increasing of diffusion coefficient of GCRs particles in comparison to the previous solar minimum and consequently the drift effect was reduced. In 2007-2008 tilt angle was relatively large (≈ 30⁰), so the drift effect might be reduced (maybe because of the weak HMF), consequently the latitudinal gradient is also smaller. At the same time coronal holes, CIRs and SW velocity were stable, so convection was strong. Due to the weak HMF diffusion was also enhanced. However, many characteristics of the heliospheric environment were not typical compared with earlier Solar Cycles, *e.g.* lower SW dynamic pressure and magnetic field strength, but high TA, together resulting in the extremely high GCRs intensities in 2009. Comparing Solar Cycles 23 and 24 it was observed (Dunzlaff *et al.*, 2008) lower effectiveness of the latitudinal transport in 2006-2009. Also (Simone *et al.*, 2011) reported smaller negative latitudinal gradient in the $qA<0$ polarity in Solar Cycle 24 comparing to previous $qA<0$, but TA was rather high (*e.g.* Heber *et al.*, 2009), so the GCRs modulation processes in the solar minimum 23/24, when $qA<0$ maybe a consequence of large latitudinal extent of HCS (Potgieter and Ferreira, 2001; Heber *et al.*, 2009). The results presented in this paper will be used in future work concerning the modeling of the transport of GCRs particles as the crucial parameters characterizing the solar modulation of GCRs by the SW in the heliosphere.

## 4. Conclusions

Using data from several neutron monitor stations of different magnetic rigidity cut-offs, we have studied the role of drift in the temporal variations of the GCRs anisotropy and its influence on the sector structure of the heliospheric magnetic field. Using the harmonic analysis method, we have analyzed the GCRs anisotropy in the $qA<0$ solar minimum period of 2007 - 2009 (Cycle 23/24) and near the $qA>0$ minimum period of 2017-2018 (Cycle 24/25). Our main results can be summarized as follows:

i) We have compared the analysis of the GCRs anisotropy using the median $R_m$ and effective $R_{ef}$ rigidity of neutron monitor response. We have found that the differences of the GCRs anisotropy parameters calculated for these two approaches: median and effective rigidity are rather negligible.
ii) The global drift due to gradient and curvature of the heliospheric magnetic field is manifested in the radial component $A_r$ of the anisotropy $A$ of GCRs. For the $qA<0$ cycle the $A_r$ is directed away from the Sun, but for the $qA>0$ magnetic cycle the $A_r$ is directed to the Sun. This type of drift effect is the source of the 22-year variation of the anisotropy of GCRs.
iii) The drift effect due to the sector structure of the heliospheric magnetic field in the anisotropy $A$ of GCRs (during the minima epochs of solar activity) is caused by the heliospheric current sheet and is visible in the azimuthal component of GCRs anisotropy.

iv) We have shown that in the solar minimum 23/24 in 2007-2009 when *qA<0*, the drift effect is not pronounced visibly in the changes of the $A_r$ component ($A_r \approx 0$), *i.e.* for this solar minimum drift effect is only ≈ 4%. Hence the diffusion dominated model of GCRs transport is more acceptable in 2007-2009. In turn, near the solar minimum 24/25 in 2017-2018 when *qA>0*, the drift effect is clearly visible and reaches ≈ 40%. So in the period of 2017-2018 the diffusion model with clearly manifested drift is acceptable.

**Acknowledgments**
Neutron monitor count rates are from Neutron Monitor Data Base: http://www01.nmdb.eu/.
The asymptotic cones of acceptance for each NM station were calculated using the FORTRAN TJ2000 program available at http://ccmc.gsfc.nasa.gov/modelweb/sun/cutoff.html.
Geomagnetic field coefficients were taken from http://www.ngdc.noaa.gov/IAGA/vmod/igrf.html.
R.M. acknowledges the financial support by the Polish National Science Centre, grant number: 2017/01/X/ST9/01023.

**Disclosure of Potential Conflicts of Interest**
The authors declare that there are no conflicts of interest.

**Publisher's Note**
Springer Nature remains neutral with regard to jurisdictional claims in published maps and institutional affiliations.

**Abbreviations**
The following abbreviations are used in this manuscript:
CC Coupling coefficients
CG Compton-Getting effect
CIR Corotating Interection Regions
GCRs Galactic Cosmic Rays
HCS Heliospheric Current Sheet
HMF Heliospheric Magnetic Field
NM Neutron Monitor
SA Solar Activity
SW Solar Wind
TA Tilt Angle.